\documentclass[conference]{IEEEtran}
\usepackage{graphicx}
\usepackage{cleveref}
\usepackage{float}

\begin{document}
\title{Top-Gated Carbon Nanotube FETs from Quantum Simulations:  using $\vec{k} \cdot \vec{p}\;$ Electronic Structure: Comparison with Experiments}

\title{Top-Gated Carbon Nanotube FETs from Quantum Simulations: Comparison with Experiments}

\author{\IEEEauthorblockN{A.~Sanchez-Soares\IEEEauthorrefmark{1}, 
T.~Kelly\IEEEauthorrefmark{1}, 
G.~Fagas\IEEEauthorrefmark{1},
J.C.~Greer\IEEEauthorrefmark{2}, and E.~Chen\IEEEauthorrefmark{3}}
\IEEEauthorblockA{\IEEEauthorrefmark{1}EOLAS Designs, Grenagh, Co. Cork, Republic of Ireland. E-mail: alfonso.sanchez@eolasdesigns.com}
\IEEEauthorblockA{\IEEEauthorrefmark{2}University of Nottingham Ningbo China, Ningbo, China.}
\IEEEauthorblockA{\IEEEauthorrefmark{3}Corporate Research, TSMC, Hsinchu, Taiwan.}}

\maketitle

\begin{abstract}
We present quantum simulations of carbon nanotube field-effect transistors (CNT-FETs) based on top-gated architectures and compare to electrical characterization on devices with 15 nm channel lengths. A non-equilibrium Green's function (NEGF) quantum transport method coupled with a $\vec{k} \cdot \vec{p}$ description of the electronic structure is demonstrated to achieve excellent agreement with the reported experimental data. Factors influencing the electrostatic control of the channel are investigated and reveal that detailed modeling of the electrostatics and the electronic band structure of the CNT is required to achieve quantitative agreement with experiment. 
\end{abstract}

% no keywords

\section{Introduction}
Recent reports for CNT-FETs indicate performance targets for sub-5 nm technologies can be met~\cite{Pitner2020}. In that work, electrical characterization and simulation for top-gated CNT-FETs is presented in which a long standing issue with nucleating high-$\kappa$ oxides on CNTs is overcome by deposition of an interfacial layer dielectric (ILX). Their CNT-FETs exhibit subthreshold swing values down to 65 mV/dec at 15 nm gate length while complying with gate leakage targets.%~\cite{Pitner2020} 

Due to difficulties in applying simple capacitance models commonly employed to describe Si CMOS operation and the need to consider various tunneling mechanisms, an approach combining detailed electrostatic modeling of the transistor geometry combined with a quantum treatment of charge transport and electronic structure are applied to determine current-voltage (IV) characteristics. Different approaches to the electrostatic modeling as well as an investigation of the electronic band structure are presented and related to subthreshold properties. The technology computer aided (TCAD) simulator $\mathcal{Q}^*$ describes in detail the properties of these transistors without the requirement for high performance computing resources or long simulation times.~\cite{Qstar}

\section{Simulation methodology}

\begin{figure}%[H]
\centering
 \includegraphics[width=0.45\textwidth]{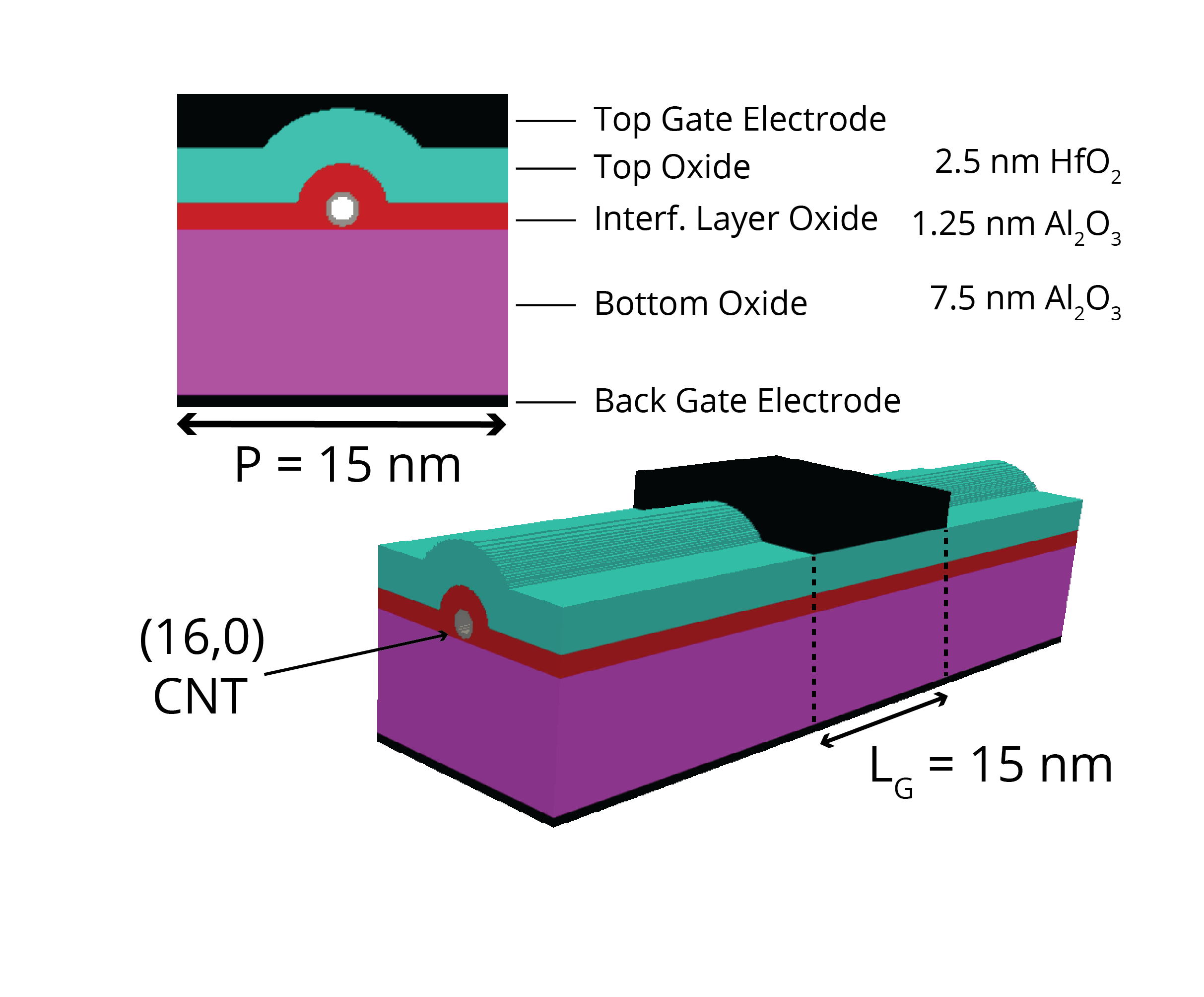}
 \caption{Geometry of simulated devices. Electrostatics are solved using periodic boundary conditions along the width direction, where a pitch of 15 nm has been simulated.}\label{fig:geo}	
\end{figure}

A coupled mode-space non-equilibrium Green function (NEGF) solver is applied employing a $\vec{k} \cdot \vec{p}$ electronic Hamiltonian capable of simulating the properties of CNT-FETs arranged in arbitrary 3D geometries. The $\mathcal{Q}^*$ software package enables fast simulations of CNT-FETs including quantum mechanical treatment of elastic and inelastic phonon scattering. We validate our method by reproducing the experimentally measured transfer characteristics of devices fabricated as described in ref. \cite{Pitner2020}; we simulate top-gated devices based on (16,0) CNTs with 15 nm gate lengths and dimensions shown in \cref{fig:geo}.

\section{Results}

\begin{figure}%[H]
\centering
 \includegraphics[width=0.45\textwidth]{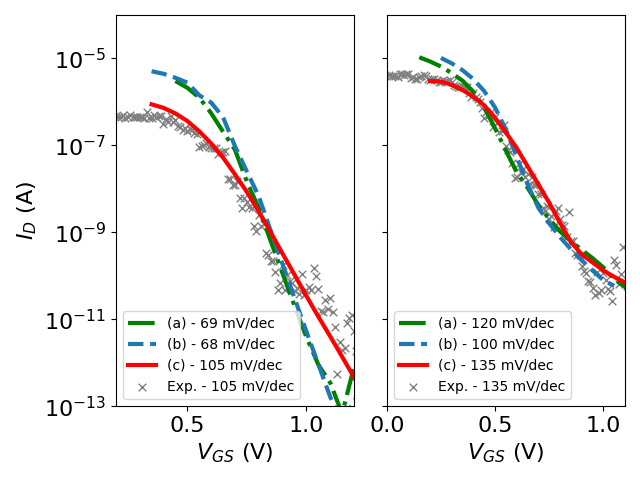}
 \caption{$I_D - V_{GS}$ characteristics of top-gated CNT-FET with 15 nm channel length at $V_{DS} = 0.05$ V (left) and $V_{DS} = 0.50$ V (right). Subthreshold swing values shown in the legend are results from fits employing current values spanning three orders of magnitude. Simulated curves have been shifted to align with experimental results.}\label{fig:iv_both}	
\end{figure}

The transfer characteristics for the fabricated devices were measured at saturation $V_{DS} = 0.50$ V and in the linear $V_{DS} = 0.05$ V regions. Comparison of experimental measurements with three simulation schemes are shown in \cref{fig:iv_both}: (a) a device without a back gate electrode having an intrinsic channel and chemically doped source/drain (S/D) extensions attached to ohmic contacts; (b) a similar device with a junctionless chemical doping scheme, and (c) a device employing electrostatic doping realised through a back-gate electrode with Schottky contacts on both ends of S/D extensions. Device (a) represents a commonly employed scheme for transistor simulations, while (b) and (c) describe configurations that are designed to increasingly match the actual experimental configuration. While all schemes are able to describe sub-$V_T$ swing degradation with increasing drain-source bias, simulations using chemical doping schemes are observed to predict better sub-$V_T$ performance. The model including electrostatic effects arising from the inclusion of back-gate electrode and metallic S/D contacts results in average sub-$V_T$ characteristics that closely track experimental results for both the linear and saturation regions. We note that although subthreshold swing values extracted from experimental data exhibit values below 80 mV/dec in some portions of the sub-$V_T$ characteristics, the values reported in \cref{fig:iv_both} correspond to averages over three orders of magnitude in current.

\begin{figure}%[H]
\centering
 \includegraphics[width=0.48\textwidth]{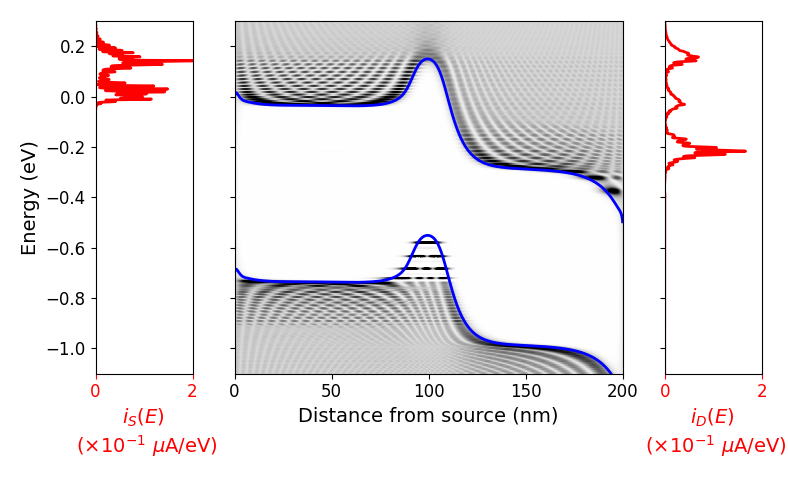}
 \caption{LDoS for a device using electrostatic doping and Schottky contacts with barrier heights of 0 eV. Zero of energy is taken to be at the source Fermi level. The red curves show the energy-resolved current at the source (left) and drain (right) side.}\label{fig:ldos_sb}	
\end{figure}

\Cref{fig:ldos_sb} shows the local density of states and energy-resolved current at both leads for a state in the subthreshold region using simulation scheme (c). Current injected into the device at the source side shows two peaks corresponding to thermionic transport and source-to-drain tunnelling; although electronic occupations are significantly larger around the source Fermi level (zero of energy in \Cref{fig:ldos_sb}), the tunnelling barrier imposed at the channel results in both peaks exhibiting similar magnitude in the steady-state solution. As carriers travel through the CNT and interact with phonon modes both peaks become broadened and a new peak emerges through the drain extension due the availability of states at lower energies. After travelling the length of the drain extension most charge carriers have participated in phonon emission processes, as current reaching the drain electrode exhibits a largest contribution at lower energies than available in the source extension.

\begin{figure}%[H]
\centering
 \includegraphics[width=0.45\textwidth]{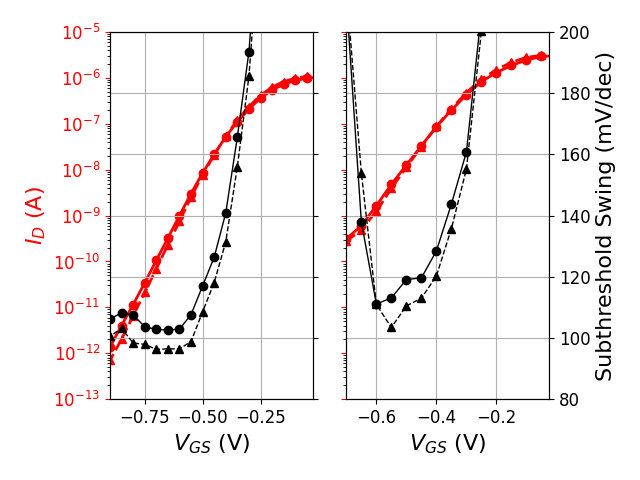}
 \caption{Subthreshold swing vs. gate bias for a back-gated device at $V_{DS} = 0.05$ V (left) and $V_{DS} = 0.50$ V (right). Reducing ILX thickness from 1.25 nm (circles) to 0.35 nm (triangles) results in a reduction of around 10 mV/dec.}\label{fig:ss_comp}	
\end{figure}

\Cref{fig:ss_comp} compares extracted sub-$V_T$ swing values at each gate bias for simulated CNT-FETs with ILX thicknesses of 1.25 nm and 0.35 nm following the gate oxide description in ref. \cite{Pitner2020}. Our model estimates that a such a reduction in ILX thickness results in an improvement of around 10 mV/dec across the entire subthreshold region.

\section{Conclusion}

CNT-FETs present excellent electrical characteristics for sub-5 nm technologies. However, details of their processing and their gate geometries require detailed electrostatic simulations to realistically capture physics behind their operation. Furthermore, features specific to the band structure of CNTs benefit from more advance treatments of the electronics structure. Phonon scattering and tunnelling mechanisms readily described within $\vec{k} \cdot \vec{p}$ enable quantitative predictions to be made at low computational effort.

To provide a validated treatment of the IV characteristics for the CNT-FET measured in ref.~\cite{Pitner2020}, a model including an explicit back gate to accurately model the electrostatics responsible for gate control of the channel was compared against more conventional simulation schemes.  Finally, investigation of the ILX thickness reveals that if the seed layer thickness for the high-$\kappa$ gate oxide can be controlled, substantial improvements in subthreshold swing can be achieved.

%\newpage

\end{document}